\documentclass[aps,prb,twocolumn,twoside,superscriptress]{revtex4-2}

\usepackage{bm}

\usepackage[version=3]{mhchem} 
\usepackage{amsmath, amssymb, amsthm, amsfonts}
\usepackage{listings, longtable, enumerate, latexsym, color, setspace}
\usepackage{dsfont, mathrsfs, array, etoolbox, bm}
\usepackage{hyperref, graphicx, subfigure, psfrag}
\usepackage{physics, braket, changes, comment, makecell}
\usepackage{mathtools}

\allowdisplaybreaks

\def\bk{{\bf k}}

\def\br{{\bf r}}

\def\bp{{\bf p}}
\def\bq{{\bf q}}
\def\bQ{{\bf Q}}

\def\be{{\bf e}}

\def\bA{{\bf A}}

\def\<{\langle}
\def\>{\rangle}

\def\bark{\bar{\bf k}}

\usepackage{stackengine}
\def\rlwd{.4pt}
\def\rlht{1.1pt}
\def\shatvrule{\rule{\rlwd}{\rlht}}
\def\shat#1{%
  \setbox0=\hbox{$#1$}%
  \stackon[0pt]{\stackon[1pt]{\ensuremath{#1}}{%
    \shatvrule\kern\wd0\kern-\rlwd\kern-\rlwd\shatvrule}}%
    {\rule{\wd0}{\rlwd}}%
}

\newcommand{\RN}[1]{%
  \textup{\uppercase\expandafter{\romannumeral#1}}%
}
\begin{document}

\newcommand{\TODO}[1]{\textcolor{red}{#1}}
\newcommand{\super}[1]{\ensuremath{^{\mathrm{#1}}}}

\title{Phonon-Assisted Radiative Lifetimes and Exciton Dynamics from First Principles}

\setcounter{page}{1}

\author{Chunhao Guo}
\affiliation{%
 Department of Materials Science and Engineering, University of Wisconsin-Madison, Madison, WI, 53706, USA
}%

\author{Gabriele Riva,\footnotemark[1]}
\thanks{CG and GR contributed equally.}
\affiliation{%
 Department of Materials Science and Engineering, University of Wisconsin-Madison, Madison, WI, 53706, USA
}%

\author{Junqing Xu}
\affiliation{Department of Chemistry and Biochemistry, University of California, Santa Cruz, CA, 95060,\footnotemark[2]}
\thanks{Current address: Department of Physics, Hefei University of Technology, 420 Feicui Road, University City, Hefei Economic and Technological Development Zone, Anhui Province, China}

\author{Jacopo Simoni}
\affiliation{%
 Department of Materials Science and Engineering, University of Wisconsin-Madison, Madison, WI, 53706, USA
}%

\author{Yuan Ping}
\email{yping3@wisc.edu}
\affiliation{%
 Department of Materials Science and Engineering, University of Wisconsin-Madison, Madison, WI, 53706, USA
}
\affiliation{Department of Physics, University of Wisconsin-Madison, Madison, WI, 53706, USA}
\affiliation{Department of Chemistry, University of Wisconsin-Madison, Madison, WI, 53706, USA}
%

\begin{abstract}
Exciton-phonon interactions play a fundamental role in phonon-assisted radiative recombination and exciton dynamics in solids. In this work, we present a first-principles framework for computing phonon-assisted radiative lifetimes and exciton dynamics at finite temperatures. Starting from the solution of the Bethe-Salpeter equation, we construct an effective excitonic Hamiltonian that incorporates both exciton-photon and exciton-phonon interactions. Phonon-assisted radiative lifetimes in anisotropic media are evaluated using time-dependent second-order perturbation theory. We further analyze the temperature and phonon-mode dependence of phonon-assisted radiative lifetime and compare our results with available experimental data. We explain the nonmonotonic temperature dependence of the phonon-assisted radiative lifetime by different mechanisms at low- and high-temperature regimes.  
Finally, we perform real-time exciton relaxation at the diagonal approximation of Lindbladian dynamics for time-resolved exciton occupation,   
providing insights into ultrafast thermalization and scattering pathways. Our \textit{ab-initio} theory offers
a detailed microscopic understanding of phonon-mediated exciton relaxation and recombination processes, and provides in-depth perspectives on phonon-assisted many-body interactions and their influence on optical properties for light-emitting and optoelectronic applications.
\end{abstract}


\maketitle

%
%

Exciton-phonon interactions play a fundamental role in the optical and electronic transport properties of semiconductors and insulators. In particular, they are essential for describing phonon-assisted radiative recombination processes and exciton dynamics in materials with indirect band gaps, where momentum conservation between the electron and hole requires the assistance of phonons during photon emission. Such phonon-mediated optical transitions are especially relevant in layered van der Waals (vdW) materials and other quantum materials, where the interplay between the many-body effects of electrons and lattice vibrations leads to rich unconventional phenomena~\cite{Von17,Rou21,brem2020phonon}. In past decades, the study of excitons, which is the quasiparticle of bound electron-hole pairs, has advanced significantly due to the development of first-principles many-body perturbation theory by solving the Bethe-Salpeter equation (BSE)~\cite{sangalli2019many,Govoni2015,deslippe2012berkeleygw, ping2013electronic}. 
It has reliably predicted exciton binding energies, exciton dispersion at finite momentum~\cite{PRLdiana}, exciton-phonon scattering rates~\cite{Lou22,chen2020exciton},  and optical spectra in bulk and low-dimensional materials~\cite{Diana2013, sangalli2019many}. 
Most of these methods have been applied in the context of first-order or direct optical transitions~\cite{FengPRB2019,palummo2015exciton,Chen19}, and do not account for second-order phonon-assisted recombination processes, which require both light-matter interaction and exciton-phonon couplings. 

Phonon-assisted luminescence, such as indirect photoluminescence (PL) observed in hexagonal boron nitride (hBN)~\cite{Von17,Rou21}, monolayer WSe\(_2\)~\cite{brem2020phonon}, or silicon~\cite{Mac58}, in principles cannot be captured by the first-order optical processes alone. Instead, second-order processes involving both exciton-photon and exciton-phonon interactions must be considered. Despite their importance, a fully \emph{ab initio} framework for phonon-assisted radiative lifetimes from many-body perturbation theory remains to be developed. Previous studies have developed second-order phonon-assisted optical transitions at the single particle level~\cite{Nof12,patrick2014unified,Tiw24} or have treated exciton-phonon and exciton-photon interactions separately~\cite{brem2020phonon,chen2020exciton,Lec23}, potentially losing coupling effects among multiple scattering channels~\cite{chan2023exciton}. In this work, we address this gap by developing a rigorous many-body formalism to compute phonon-assisted radiative recombination and time-resolved exciton dynamics from first principles. Our methodology is based on the construction of an effective exciton-photon-phonon Hamiltonian in second quantization, incorporating the excitonic states from the solution of the finite-momentum BSE~\cite{sangalli2019many} and the exciton-phonon coupling derived from electron-phonon matrix elements~\cite{Giu17,Ant22,Pal22,chen2020exciton}. We evaluate the phonon-assisted radiative rates using time-dependent second-order perturbation theory~\cite{sakurai2020modern}, explicitly accounting for phonon emission and absorption channels. Furthermore, to capture the out-of-equilibrium dynamics of excitons in open quantum system, we derive a quantum master equation in the Lindblad form~\cite{Ros14,Ros15,xu2020spin, Xu2024} that describes the evolution of the exciton density matrix under phonon scatterings. In the semi-classical limit, this reduces to the Boltzmann transport equation (BTE)~\cite{rosati2020temporal,Kat23,Chen22}, enabling real-time simulation of exciton relaxation and thermalization.


\begin{figure}
    \includegraphics[width=0.25\textwidth]{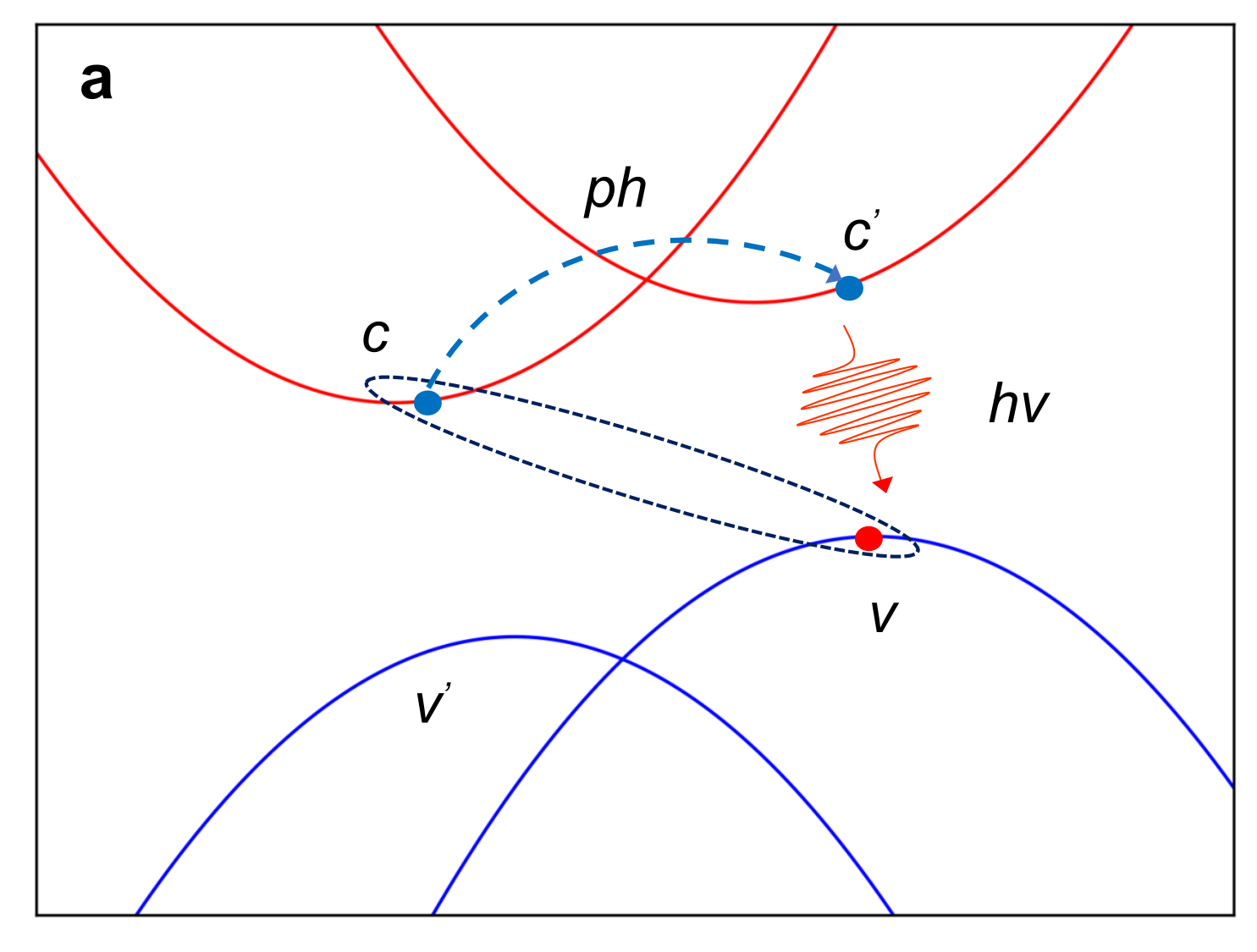}%
    \includegraphics[width=0.25\textwidth]{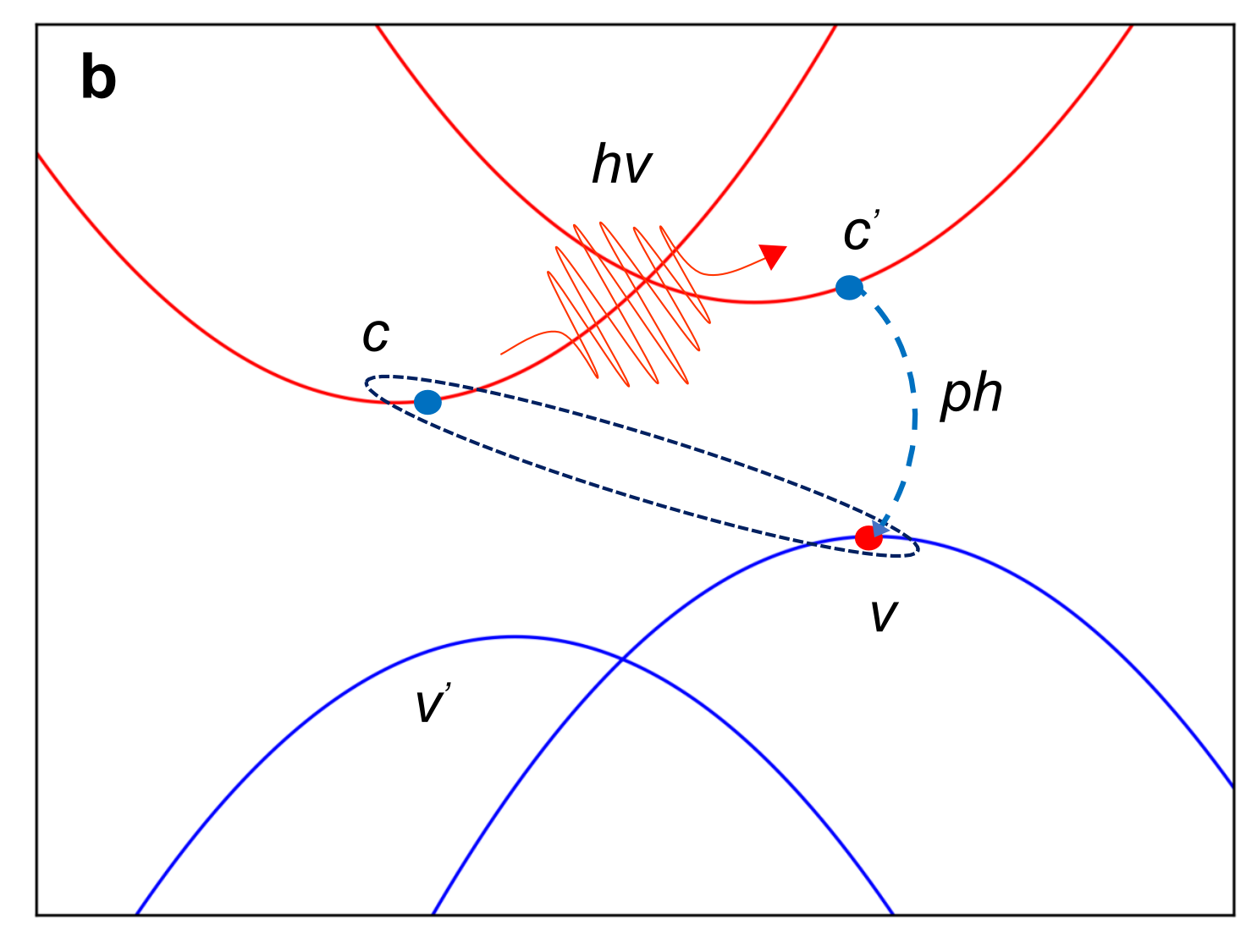}%
    \centering
\caption{\label{fig-schemetic} Schematic illustration of phonon-assisted radiative processes. (a) Intraband phonon scattering (ph) combined with interband photon emission ($h\nu$), and (b) interband phonon scattering combined with intraband photon emission. 
}

\end{figure}

We first derive the effective exciton-photon-phonon Hamiltonian starting from the one-particle light-matter interaction (electron-photon) Hamiltonian and the electron-phonon Hamiltonian. The exciton, as the eigenstate of the BSE Hamiltonian, is a quasiparticle defined in the electron-hole space~\cite{Oni02}. The exciton creation operator can be expressed in the form \(\hat{c}^\dagger_{S(\bQ)} = \sum_{vc\bk} A^{S(\bQ)}_{vc\bk} \hat{d}^\dagger_{c\bk+\bQ} \hat{d}_{v\bk},\) where \(\hat{c}^\dagger_{S(\bQ)}\) creates an exciton in state \(S\) with momentum \(\bQ\), \(\hat{d}^\dagger_{c\bk+\bQ}\) and \(\hat{d}_{v\bk}\) are electron creation and annihilation operators, respectively. \(A^{S(\bQ)}_{vc\bk}\) is the exciton amplitude. This formalism implies that a single exciton corresponds to an interband transition. In contrast, describing an intraband transition requires the involvement of two excitons. We start with the non-interacting Hamiltonian, which includes the exciton, photon, and phonon fields:

\begin{align}\label{H_0:eq}
    \hat{H}_{\rm 0} =& \hat{H}_{\rm ex} + \hat{H}_{\rm ph} + \hat{H}_{\rm \gamma} \nonumber \\ 
    =& \sum_{S,\bQ} \hbar \omega_{S(\bQ)} \hat{c}^\dagger_{S(\bQ)} \hat{c}_{S(\bQ)} 
    + \sum_{\nu \bq} \hbar \omega_{\nu \bq} \left( \hat{b}^\dagger_{\nu \bq} \hat{b}_{\nu \bq} + \frac{1}{2} \right) \nonumber
    \\+& \sum_{\lambda \bark} \hbar \omega_{\lambda \bark} \left( \hat{a}^\dagger_{\lambda \bark} \hat{a}_{\lambda \bark} + \frac{1}{2} \right).
\end{align}
where \(\nu\), and \(\lambda\) are indices representing phonon, and photon modes, respectively, and \(\bq\), and \(\bark\) denote their corresponding wavevectors. The field operators \(\hat{b}^{(\dagger)}, \hat{a}^{(\dagger)}\) represent the annihilation (creation) operators for phonons, and photons, respectively.
Using the plane wave representation of the vector potential \(\bA\), extending the electron-photon Hamiltonian~\cite{ger07} \(\hat{H}^{\rm e-\gamma}=-\frac{e}{m} \hat{\bA}  \cdot \hat{\bp}\) to the excitonic basis \(\hat{H}^{\rm ex-\gamma}=-\frac{e}{m_{\rm ex}} \hat{\bA} \cdot \hat{\bp}_{\rm ex}\), the exciton-photon Hamiltonian is written as:

\begin{align}\label{ex-photonH_def:eq}
    \hat{H}^{\rm ex-\gamma} =& i \sum_{\lambda \bark} \sqrt{\frac{e^2}{2\hbar \omega_{\lambda \bark} V \epsilon_0}} \nonumber \\&
    \left( \be_{\lambda \bark} \hat{a}_{\lambda \bark} e^{i\bark \cdot \br} 
    + \be^*_{\lambda \bark} \hat{a}^\dagger_{\lambda \bark} e^{-i\bark \cdot \br} \right) 
    \cdot [\hat{\br}, \hat{H}_{\rm ex}],
\end{align}
where \([\hat{\br}, \hat{H}_{\rm ex}]\) represents the commutator between the position operator \(\hat{\br}\) and the excitonic Hamiltonian \(\hat{H}_{\rm ex}\). Considering the electron-phonon Hamiltonian at first order in atomic displacement~\cite{Giu17}, it can be expressed as: \(\hat{H}^{\rm e-ph} = \sum_{\bk,\bq} \sum_{mn\nu} g_{mn\nu}(\bk,\bq) \hat{d}^\dagger_{m,\bk+\bq} \hat{d}_{n,\bk} (\hat{b}_{\bq \nu} + \hat{b}^\dagger_{-\bq \nu}),\) where \(\hat{d}\) is the electron field operator, and \(m\), \(n\) represent the band indices. The term \(g_{mn\nu}(\bk,\bq) = \langle m\bk+\bq | \Delta_{\bq\nu} v^{\rm KS} | n\bk \rangle\) denotes the electron-phonon coupling matrix in the mode representation, with \(\Delta_{\bq\nu} v^{\rm KS}\) being the derivative of the Kohn-Sham potential from DFPT calculations. The Born–von Karman (BvK) boundary condition is applied to plane waves~\cite{Giu17}. Consequently, the summation over wavevectors is restricted to the first Brillouin zone, and the normalization factor \(N_p^{-1/2}\) is omitted for simplicity. Therefore, the total Hamiltonian \(\hat{H}\) is defined as \(\hat{H}_{\rm 0} + \hat{H}_{\rm int}\), where \(\hat{H}_{\rm int}\) represents the interacting Hamiltonian, given by: \( \hat{H}_{\rm int} = \hat{H}^{\rm ex-\gamma} + \hat{H}^{\rm ex-ph}_{\rm intra} + \hat{H}^{\rm ex-ph}_{\rm inter},\)

\begin{equation}\label{H_intra_def:eq}
    \hat H^{\rm ex-ph}_{\rm intra}= \sum_{SS'}\sum_{\bq \bQ \nu} \mathcal{G}_{S'S\nu}(\bQ,\bq) \hat c^{\dagger}_{S'(\bQ+\bq)}\hat c_{S(\bQ)} (\hat b_{\bq \nu}+\hat b^{\dagger}_{-\bq \nu}) \nonumber;  
\end{equation}
\begin{align}\label{H_inter_def:eq}
    \hat{H}^{\rm ex-ph}_{\rm inter} &= \sum_{S\bQ\nu} \mathcal{G}_{S\nu}(\bQ) \hat{c}_{S(\bQ)} (\hat{b}_{-\bQ \nu} + \hat{b}^\dagger_{\bQ \nu}),
\end{align}
the intraband and interband exciton-phonon interaction Hamiltonians are derived by projecting the electron-phonon coupling from the one-particle basis onto the excitonic basis, where $\mathcal{G}$ is the interband and intraband exciton-phonon matrix element:

\begin{equation}\label{G_intra:eq}
\begin{split}
    \mathcal{G}_{S^{\prime}S\nu} (\mathbf{Q},\mathbf{q}) =& \sum_{vcc'\mathbf{k}}A^{S^{\prime} (\mathbf{Q+q})*}_{vc\mathbf{k}} A^{S (\mathbf{Q})}_{vc' \mathbf{k}}  g_{cc' \nu} (\mathbf{k+Q,q}) \nonumber \\-& \sum_{cvv^{\prime}\mathbf{k}}A^{S' (\mathbf{Q+q})*}_{vc\mathbf{k}} A^{S (\mathbf{Q})}_{v'c \mathbf{k+q}} g_{v'v \nu} (\bk,\bq); \nonumber
\end{split}
\end{equation}
\begin{equation}\label{G_inter:eq}
\begin{split}
    \mathcal{G_{S\nu}(\bQ)}=\sum_{vc\bk}A^{S(\bQ)}_{vc \bk} g_{vc\nu}(\bk+\bQ,-\bQ).
\end{split}
\end{equation}
The phonon-assisted radiative transition probability is expressed in terms of the time-dependent perturbative coefficients~\cite{sakurai2020modern} \(c^{(n)}_{i \to f}(t)\) as: \(P_{i \to f}(t) = \abs{c^{(1)}_{i \to f}(t) + c^{(2)}_{i \to f}(t) + \cdots}^2,\) for the specific \(i \to f\) transition under consideration, the first-order coefficient \(c^{(1)}_{i \to f}(t)\) vanishes. Contributions from higher-order terms beyond the second order are not included in this analysis. The transition rate includes contributions from both interband and intraband processes:
\begin{widetext}
\begin{align}\label{rate:eq_12}
    \gamma_{S(\bQ)}&= \frac{2\pi}{\hbar^4} \sum_{\lambda,\bark,\bq,\nu,\pm}    \left(n_{\nu,\mp \bq}+\frac{1}{2}\pm \frac{1}{2}\right)\delta(\omega_{\lambda,\bark}\pm \omega_{\nu,\mp \bq}-\omega_{S(\bQ)}) \nonumber \\
    & \lim_{\eta\to 0^+}\left| \sum_{S'} \left[ \frac{\mathcal{G_{S'\nu}(-\bq)} \mathcal{M}_{S'(-\bq)S(\bQ),\lambda\bark}}{\omega_{S'(-\bq)}+ \omega_{\lambda, \bark}-\omega_{S(\bQ)}-i\eta} + \frac{\mathcal{G}_{S'S\nu}(\bQ,\bq)\mathcal{M}_{S'(\bQ+\bq),\lambda\bark}}{\omega_{S'(\bQ+\bq)}\pm \omega_{\nu,\mp \bq}-\omega_{S(\bQ)}-i\eta} \right] \delta_{\bQ+\bq,-\bark}\right|^2 ,
\end{align}
\end{widetext}
where the detailed derivation of second-order time-dependent perturbation theory is provided in the SI Sec.~S1-S3. The expression of interband and intraband exciton-photon matrix elements \(\mathcal{M}\), within the dipole approximation, is provided as:

\begin{equation}\label{M_intra:eq}
\begin{split}
    \mathcal{M}_{S'(\bQ')S(\bQ),\lambda\bark}=i\hbar\sqrt{\frac{e^2}{2\hbar\omega_{\lambda \bark}\epsilon_0 V}} &(\omega_{S(\bQ)}-\omega_{S'(\bQ')}) \be^*_{\lambda \bark} \nonumber \\&\cdot \langle S'(\bQ')|\hat{\br} |S(\bQ)\rangle; \nonumber
\end{split}
\end{equation}
\begin{equation}\label{M_inter:eq}
    \mathcal{M}_{S(\bQ),\lambda\bk}=i\hbar\sqrt{\frac{e^2}{2\hbar\omega_{\lambda \bk}\epsilon_0 V}} \omega_{S(\bQ)}\be^*_{\lambda, \bk}\cdot \bra{G} \hat{\br} \ket{S(\bQ)}.
\end{equation}

In the following discussion, we focus on the phonon-assisted radiative process, specifically, intraband exciton-phonon coupling, which is described by the second term in Eq.~\eqref{rate:eq_12} and depicted in Fig.~\ref{fig-schemetic}(a). The interband exciton-photon coupling, which is described by the first term in Eq.~\eqref{rate:eq_12} and depicted in Fig.~\ref{fig-schemetic}(b), corresponding to 
%
%
transitions between the valence and conduction bands induced by phonon excitation, is typically negligible in semiconductors, because of the large energy gap between these bands compared to phonon energies. 

Considering the propagation of light in uniform non-magnetic anisotropic media~\cite{jackson1999classical, yang2014fundamentals}, the relative dielectric tensor $\stackrel{\leftrightarrow}{\epsilon}$ can be diagonalized along the principal optical axes.
In our later discussion, we focus on a special case when $\epsilon_{xx} = \epsilon_{yy} = \epsilon_{\parallel}$ and $\epsilon_{zz} = \epsilon_{\perp}$, which is sufficient to describe out-of-plane anisotropic medium. We convert the sum of the photon wavevector \(\bark\) into an integral \(\frac{V}{(2\pi)^3} \int_\Omega d^3\bark\) in Eq.(\ref{rate:eq_12}), then use the dispersion relation \(\omega_{\lambda \bark} = c \abs{\bark} /n_{\lambda}\), where \(n_{\lambda}\) is the refractive index and \(\lambda = \pm 1\) represents the photon mode. We finally obtain the state-resolved phonon-assisted radiative rate, \(\gamma_{S(\bQ)}\) as follows (the details can be found in the SI Sec.~S4):
\begin{align}\label{eq:gamma_Q}
    \gamma_{S(\bQ)} = \frac{4}{3} \alpha \sum_{\nu \bq \pm} \left( \omega_{S(\bQ)} \mp \omega_{\nu, \mp\bq} \right) \abs{ \stackrel{\leftrightarrow}{\mathcal{E}}^{\frac{1}{2}} \boldsymbol{\Lambda}_{S\nu\pm}(\bQ,\bq)}^2 n^{\pm}_{\nu,\bq},
\end{align}
where the dimensionless fine-structure constant \(\alpha\) is expressed as \(\alpha = \frac{e^2}{4\pi\epsilon_0\hbar c}\), the dimensionless effective amplitude \(\boldsymbol{\Lambda}\) in the long wave-length limit is expressed as: \(\boldsymbol{\Lambda}_{S\nu\pm}(\bQ,\bq) \nonumber = \frac{1}{\hbar c} \sum_{S^{\prime}} \omega_{S^{\prime}, \mathbf{0}} \frac{\boldsymbol{\mu}_{S^{\prime}(\mathbf{0})}\mathcal{G}_{S^{\prime}S\nu}(\bQ,\bq)}{\omega_{S^{\prime}, \mathbf{0}} - \omega_{S, \bQ} \pm \omega_{\nu, \mp\bq} - i\eta} \delta_{\bQ, -\bq},\) the \(\boldsymbol{\mu}\) represents the exciton transition dipole moment: \(\boldsymbol{\mu}_{S^{\prime}(\bQ^{\prime})} = \bra{G} \hat{\mathbf{r}} \ket{S^{\prime}(\bQ^{\prime})}\). 
The phonon occupation factor for phonon emission (+)/absorption (-) process is \(n^{\pm}_{\nu,\bq} = n_{\nu,\mp \bq}+\frac{1}{2}\pm \frac{1}{2}\). The dimensionless effective tensor is \(\stackrel{\leftrightarrow}{\mathcal{E}}\) is expressed as:

\begin{align}
\stackrel{\leftrightarrow}{\mathcal{E}} = 
\begin{pmatrix}
\frac{\epsilon_{\parallel}^{-1/2} \epsilon_{\perp} + 3 \epsilon_{\parallel}^{1/2}}{4} & 0 & 0 \\
0 & \frac{\epsilon_{\parallel}^{-1/2} \epsilon_{\perp} + 3 \epsilon_{\parallel}^{1/2}}{4} & 0 \\
0 & 0 & \epsilon_{\parallel}^{1/2}
\end{pmatrix}.
\end{align}
The finite-temperature phonon-assisted radiative rate \(\gamma\) is calculated as a thermal average of \(\gamma_{S(\bQ)}\) in the exciton dilute limit: \(\gamma = \braket{\gamma_{S(\bQ)}} = \frac{1}{\mathcal{Z}} \sum_{S\bQ} e^{-\beta\hbar\omega_{S(\bQ)}} \gamma_{S(\bQ)},\) where the partition function is given by \(\mathcal{Z} = \sum_{S\bQ} e^{-\beta\hbar\omega_{S(\bQ)}}\)~\cite{wu2019dimensionality}. The finite-temperature phonon-assisted radiative lifetime \(\tau\) is defined as the inverse of the phonon-assisted radiative rate.

To study the exciton dynamics and thermalization, we derived the quantum master equation in the Lindblad form~\cite{Xu2024} for exciton-phonon dynamics (the details can be found in the SI Sec.~S5):
\begin{align}
\frac{\mathrm{d}\boldsymbol{\rho}_{12}}{\mathrm{d}t}= & \frac{1}{2}\sum_{345}\left(\left(\mathds{1}+\boldsymbol{\rho}\right)_{14}\mathcal{P}_{4253}\boldsymbol{\rho}_{53}-\left(\mathds{1}+\boldsymbol{\rho}\right)_{34}\mathcal{P}_{3415}^{*}\boldsymbol{\rho}_{52}\right) \nonumber \\&+h.c.,
\end{align}
where the scattering terms are defined as: \(\mathcal{P}_{1234} = \sum_{\bq\nu\pm}\mathcal{A}_{13}^{\bq\nu\pm}\mathcal{A}_{24}^{\bq\nu\pm,*}\) with \(\mathcal{A}_{13}^{\bq\nu\pm} = \mathcal{D}_{13}^{\bq\nu\pm}\sqrt{n_{\nu, \mp\bq} + \frac{1}{2} \pm \frac{1}{2}}\). The exciton-phonon coupling term \(\mathcal{D}_{12}^{\bq\nu\pm}\) is expressed as: \(
\mathcal{D}_{12}^{\bq\nu\pm} = \sqrt{\frac{2\pi}{\hbar^2}} \mathcal{G}_{12}^{\bq\nu}\delta^{\frac{1}{2}}_{\rm G}\left(\omega_{1} - \omega_{2} \pm \omega_{\mp\bq\nu}\right),\) where \(\delta_{\text{G}}\) represents an energy-conserving Dirac delta function broadened
to a Gaussian. Here, the exciton-phonon matrix element \(\mathcal{G}_{S^{\prime}S\nu}(\mathbf{Q},\mathbf{q})\) is denoted for simplicity as \(\mathcal{G}_{12}^{\bq\nu}\), where the indices are defined as \(1 \coloneqq S^{\prime}(\bQ + \bq)\) and \(2 \coloneqq S(\bQ)\). \( \boldsymbol{\rho}\) represents the exciton density matrix. Notably, the primary distinction between the Lindblad equation for the exciton dynamics and that for the electron dynamics\cite{xu2020spin, Xu2024, PhysRevB.104.184418} lies in the term \(\mathds{1} - \boldsymbol{\rho}\) being replaced by \(\mathds{1} + \boldsymbol{\rho}\). This difference arises due to the distinct (anti)commutation relations governing fermionic and bosonic operators. The related detailed derivation can be found in the SI Sec.~S5. In this study, we focus on exciton occupation changes due to exciton-phonon scatterings, where the phase coherence information between different exciton states (corresponding to the off-diagonal elements of exciton density matrices) is not important, which will leave for our future studies. Therefore, by applying the diagonal (semiclassical) approximation to exciton density matrices and Lindbladian operators, 
we obtain the (semiclassical) quantum master equation:
\begin{align}\label{eq:ex-ph-master}
\frac{\mathrm{d}}{\mathrm{d}t}\mathcal{N}_{2} =& \sum_{1} \Gamma_{21} \mathcal{N}_{1} \left(1 + \mathcal{N}_{2}\right) - \sum_{1} \Gamma_{12} \mathcal{N}_{2} \left(1 + \mathcal{N}_{1}\right),
\end{align}
where the scattering rate \(\Gamma_{12}\) describes the transition rate between exciton states \(1,2\). This rate is expressed as:
\begin{align}\label{eq:bte}
\Gamma_{12} =& \mathcal{P}_{1122} = \sum_{\bq\nu\pm}\mathcal{A}_{12}^{\bq\nu\pm}\mathcal{A}_{12}^{\bq\nu\pm,*} \nonumber \\=& \frac{2\pi}{\hbar^2} \sum_{\bq\nu\pm} \abs{\mathcal{G}^{\bq\nu}_{12}}^2 \left(n_{\nu, \mp\bq} + \frac{1}{2} \pm \frac{1}{2}\right) \delta_{\rm G}(\omega_{1} - \omega_{2} \pm \omega_{\mp\bq\nu}),
\end{align}
which aligns with the transition rate derived from the Fermi's golden rule. 


\begin{figure}
    \includegraphics[width=0.25\textwidth]{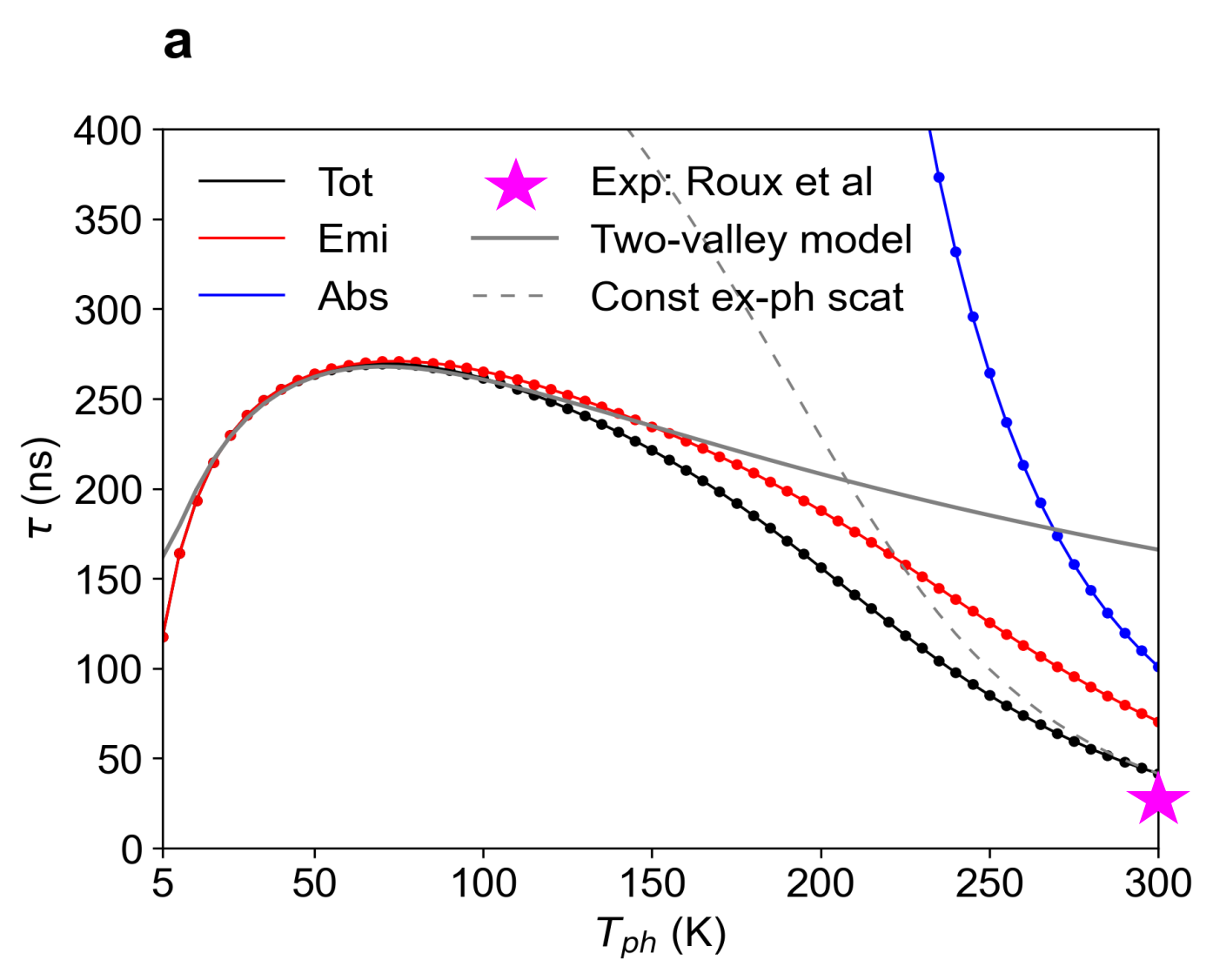}%
    \includegraphics[width=0.25\textwidth]{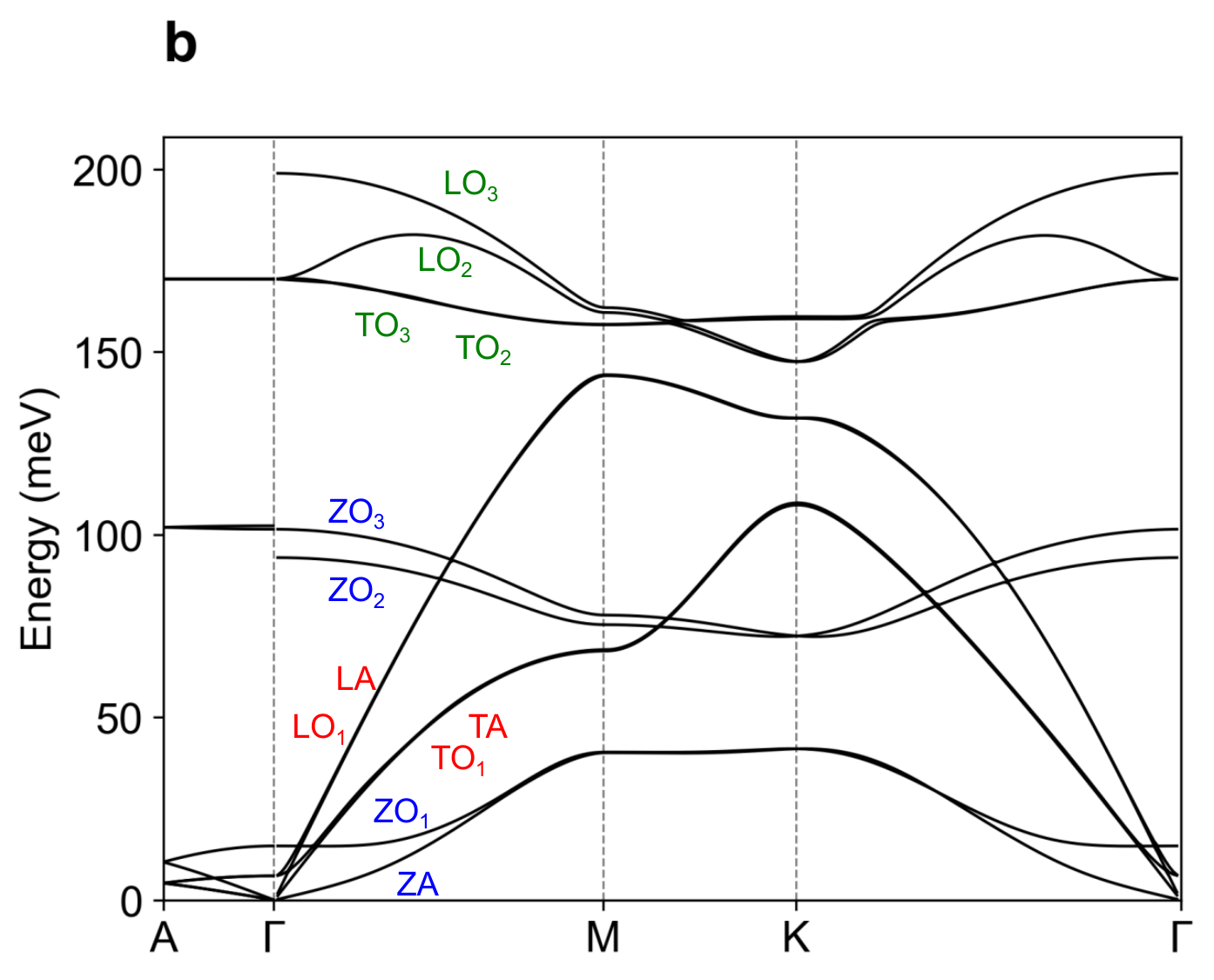}\\
    \includegraphics[width=0.25\textwidth]{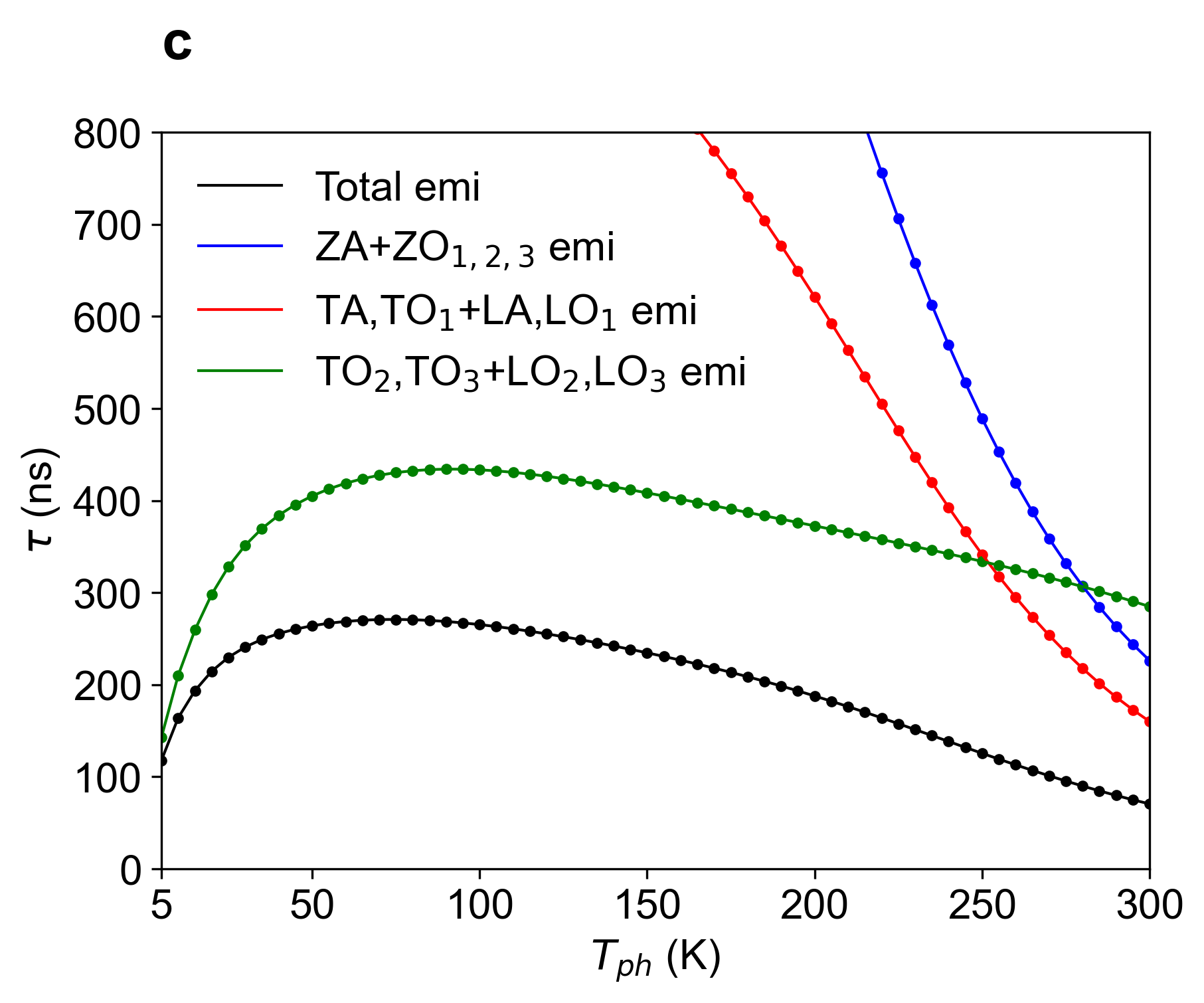}%
    \includegraphics[width=0.25\textwidth]{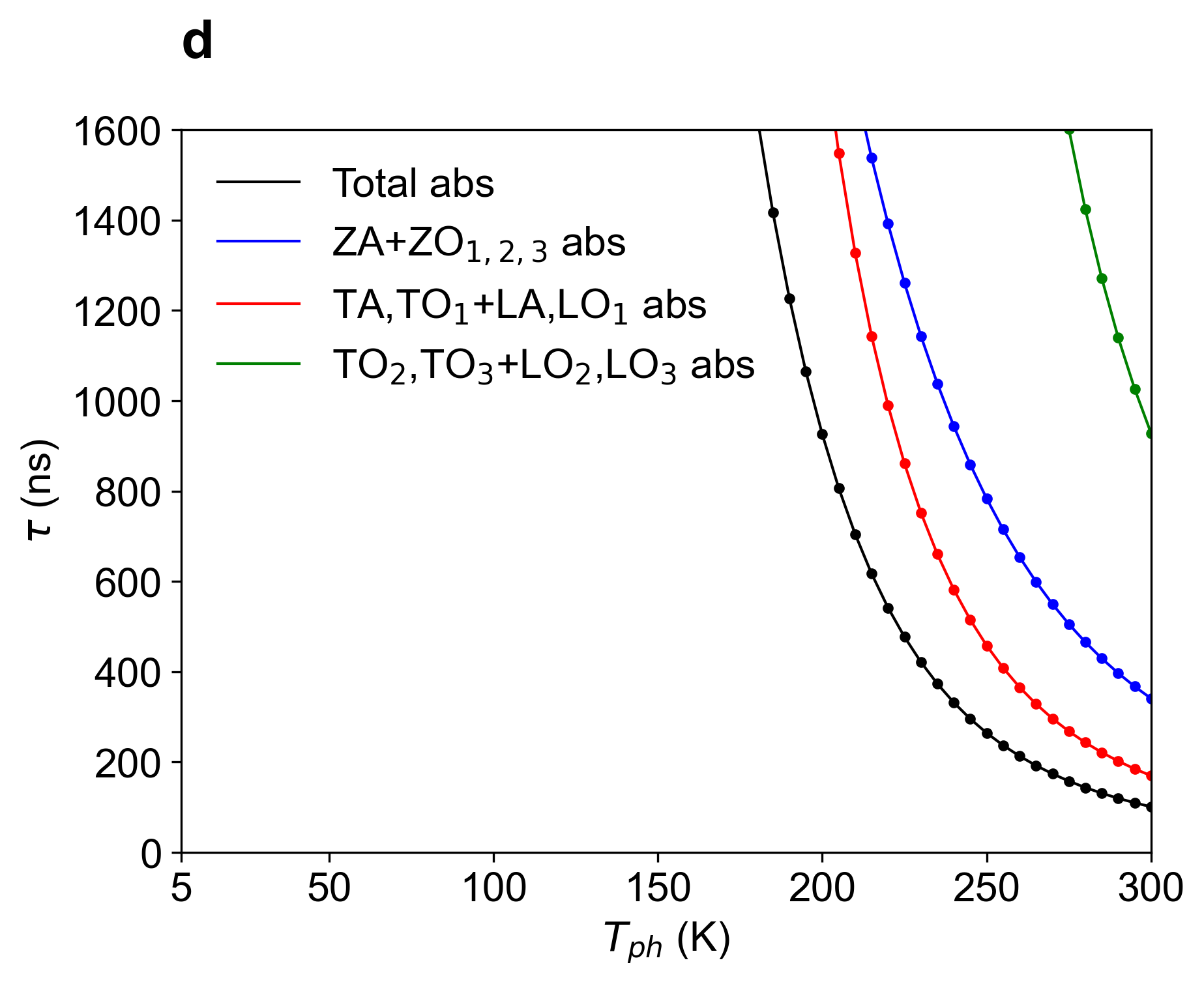}%
    \centering
\caption{\label{fig-rad-lifetime} 
Temperature dependence of phonon-assisted radiative lifetime in bulk hBN. 
(a) Total radiative lifetime (black), with contributions from phonon emission (red) and phonon absorption (blue) shown separately. The magenta star marks the experimental radiative lifetime of 27 ns at room temperature, reported by S.~Roux et al.~\cite{Rou21}. The gray solid line represents a fit using a two-valley model, capturing the non-monotonic behavior at low temperatures. The gray dashed line corresponds to Eq.~\ref{eq:gamma_Q} using constant exciton-phonon coupling \(\mathcal{G}\).
(b) Labeling of phonon branches in the phonon band structure of bulk hBN. 
(c) Phonon mode contributions to the radiative lifetime from phonon emission processes. 
(d) Phonon mode contributions to the radiative lifetime from phonon absorption processes.}
\end{figure}

\begin{figure}
    \includegraphics[width=0.25\textwidth]{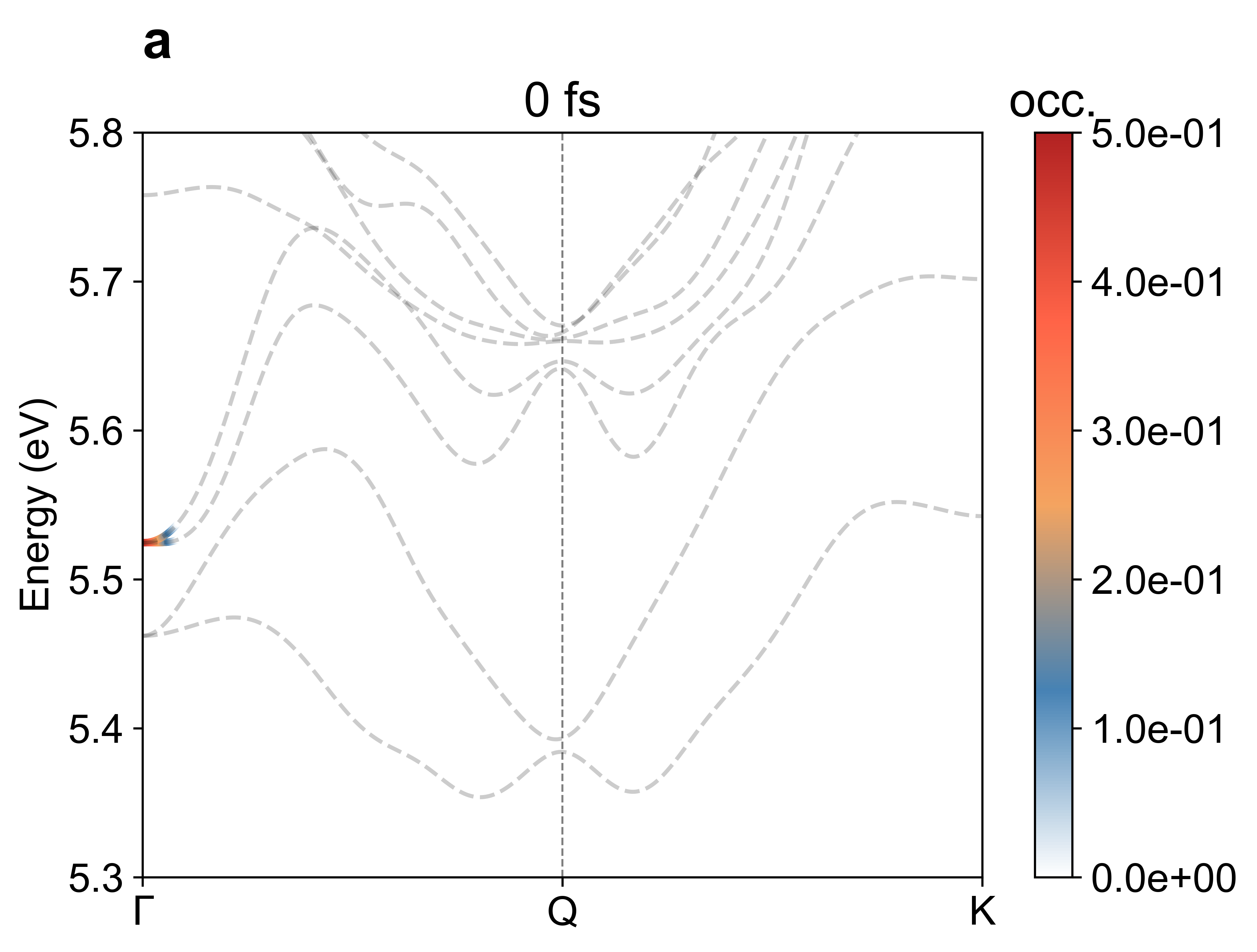}%
    \includegraphics[width=0.25\textwidth]{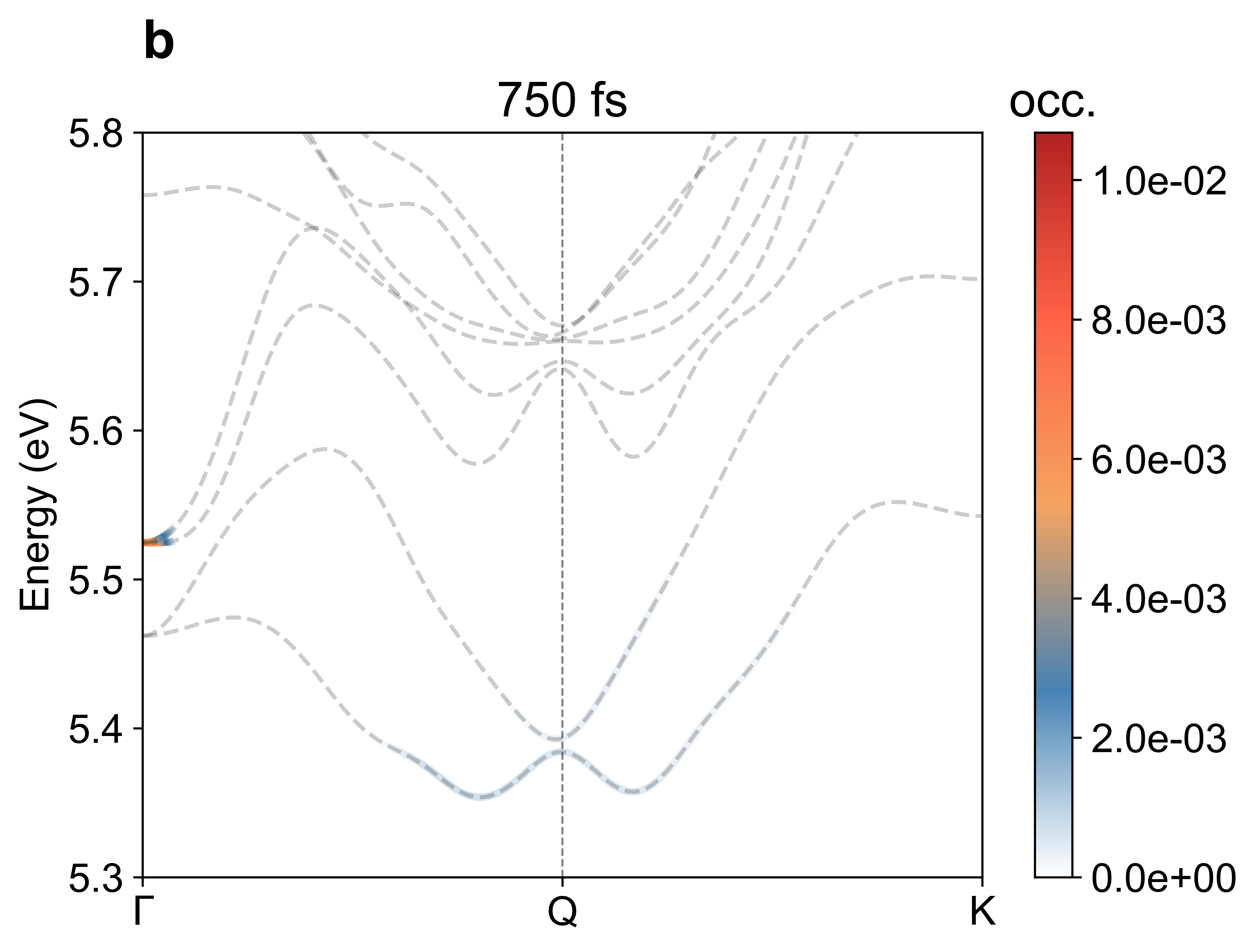}\\
    \includegraphics[width=0.25\textwidth]{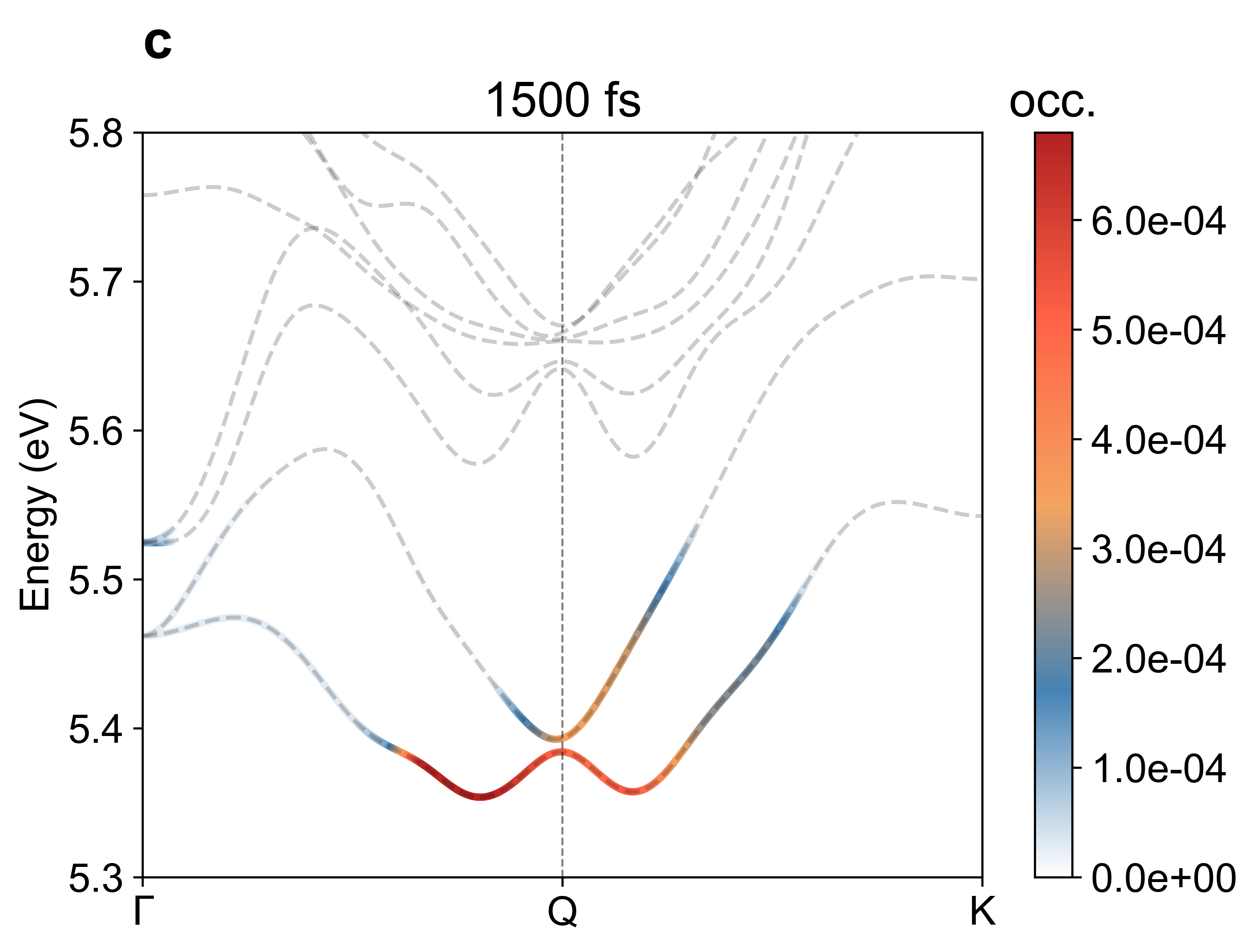}%
    \includegraphics[width=0.25\textwidth]{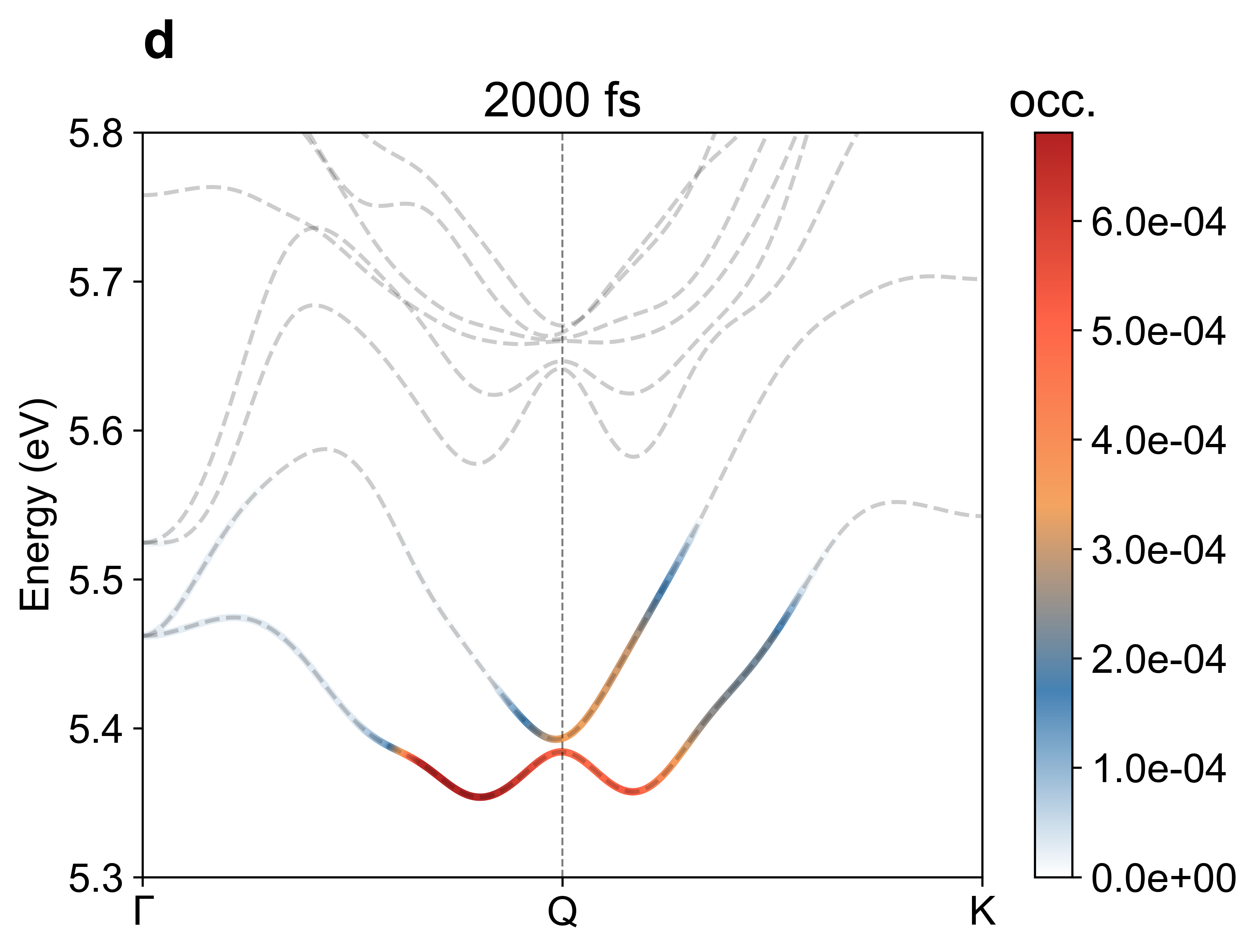}
    \centering
\caption{\label{fig-rt-occ} 
Real-time exciton occupation of bulk hBN at room temperature simulated using the real-time semi-classical quantum master equation Eq.~(\ref{eq:ex-ph-master}) at different time steps: 
(a) 0 fs (initial step), 
(b) 750 fs, 
(c) 1500 fs, and 
(d) 2000 fs (final step). 
Exciton occupation is projected onto the exciton band structure and plotted as a colormap along \(\Gamma \to Q \to K.\)
}

\end{figure}

Next, we present the results of temperature dependent phonon-assisted radiative lifetime in Fig.~\ref{fig-rad-lifetime}(a). The black curve shows the total radiative lifetime, which includes both phonon emission (the red curve) and phonon absorption contributions (the blue curve). At temperature below 150K, the total radiative lifetime exhibits a non-monotonic behavior, which can be understood qualitatively by a two-valley model. 
We approximate the exciton energy surface as a two-level model consisting of valleys 1 and 2, where valley 1 dominates the radiative process. Neglecting exciton dispersion, we assume the radiative rate for valley 1 is given by \(\gamma_1(T) = \Tilde{\gamma}_1 (n_{\text{ph}} + 1)\), where \(n_{\text{ph}} = (e^{\beta \hbar \Tilde{\omega}} - 1)^{-1}\) is the Bose-Einstein phonon occupation factor for a single effective phonon mode of frequency \(\Tilde{\omega}\). 
At low temperatures, the phonon emission process dominates. The total low temperature thermal-averaged radiative rate is then given by:
\begin{equation}\label{eq:model-low-temp}
\gamma(T) \sim \Tilde{\gamma}_1 \left( \frac{1}{1 + e^{-\beta \Delta E}} \right) \left( \frac{1}{e^{\beta \hbar \Tilde{\omega}} - 1} + 1 \right),
\end{equation}
where the first bracket term accounts for the thermal occupation of the dominate valley and decreases with increasing temperature, while the second bracket term captures the temperature dependence of phonon emission occupation factor and increases with temperature. Their competition can lead to the observed nonmonotonic behavior in the total and phonon emission radiative lifetimes at low temperature. 
We fitted this two-valley model to phonon-assisted radiative rate $\gamma(T)$ from first-principles calculations. 
A detailed discussion of this two-valley model and effective parameters fit in Eq.~\ref{eq:model-low-temp} is provided in the SI Sec.~S6. 
The resulting curve, shown as the gray solid line in Fig.~\ref{fig-rad-lifetime}(a), agrees with the first-principles results for the total and phonon-emission radiative lifetimes in the low-temperature regime with the non-monotonic behavior. At temperature above 150K, the radiative lifetime exhibits a monotonic decrease with increasing temperature. This behavior is mainly driven by the sharp increase in the phonon population as the thermal energy \(k_{\text{B}}T\) becomes comparable to the phonon energy. To analyze this effect, we set the exciton-phonon matrix elements in Eq.~\ref{eq:gamma_Q} to a constant value that matches the calculated radiative lifetime at room temperature. The resulting curve, shown as the gray dashed line in Fig.~\ref{fig-rad-lifetime}(a), reproduces the decreasing trend observed in the full first-principles calculation. This comparison indicates that, unlike the low-temperature regime, where differences in valley-dependent scattering rates dominate, the high-temperature behavior is governed by the thermal-induced increase of the phonon population and is less dependent on the detailed exciton-phonon coupling structure.

We analyze the phonon-mode contributions to the phonon-assisted radiative lifetime, separating the effects of phonon emission and absorption, as shown in Fig.~\ref{fig-rad-lifetime}(b-d). 
For the phonon emission process in Fig.~\ref{fig-rad-lifetime}(c), high-energy optical modes (green) dominate at low temperatures due to their strong exciton-phonon coupling strength. The phonon-mode resolved exciton-phonon matrix elements are provided in the SI Sec.~S7.
Their contributions are relatively less sensitive to temperature due to much higher phonon energies compared to thermal energy \(k_{\text{B}}T\). As the temperature increases, low-energy optical and acoustic modes (red and blue) become increasingly important, as their phonon occupations rise rapidly when \(k_{\text{B}}T\) becomes comparable to their phonon energies. In contrast, the phonon absorption process in Fig.~\ref{fig-rad-lifetime}(d) shows negligible contributions from all phonon modes at low temperatures due to the lack of thermal population of phonons. As the temperature increases, low-energy optical and acoustic modes become dominant for phonon absorption processes, and correspondingly the radiative rate, as their phonon occupations rise significantly. 

We finally study exciton dynamics based on Eq.~(\ref{eq:ex-ph-master}), the computational details can be found in SI Sec.~S6-S7. 
Our analysis of exciton occupation at different time steps in Fig.~\ref{fig-rt-occ}(a-d) reveals the exciton redistribution (thermalization) toward finite-momentum energy valleys close to Q. Initially, the exciton population is prepared at the first bright optical exciton state as shown in Fig.~\ref{fig-rt-occ}(a). We set the exciton occupation number to \(1\), consistent with the exciton dilute limit. As the system evolves, exciton-phonon scattering leads to relaxation
and redistribution towards lower-energy states. Our simulations indicate that the relaxation process towards the energy minimum follows an exponential decay law, see SI Sec.~S6, with a fitted valley relaxation time of approximately \(114\)~fs at room temperature. This timescale is consistent with experimental PL linewidth measurements at room temperature~\cite{Von17}, which report a linewidth \(\Delta\) of around \(30\)~meV (corresponding to an effective relaxation time \(\tau \approx 60\)~fs, based on the optical theorem relation \(\tau^{-1} = \frac{2}{\hbar}\Delta\)).
Note that our exciton Lindbladian dynamics formalism is well suited for extending beyond the semiclassical Boltzmann equation by incorporating exciton density matrix and full Lindbladian operators. 
This framework provides a direct pathway for studying exciton dephasing and exciton spin dynamics, capturing quantum coherence effects~\cite{Xu2024-lw,PhysRevB.111.115113} that are absent in the semiclassical limits. 

In summary, we developed a comprehensive \emph{ab initio} framework for evaluating phonon-assisted radiative processes and exciton dynamics in semiconductors, based on time-dependent second-order perturbation theory and many-body perturbation theory, starting with an effective exciton-photon-phonon Hamiltonian. 
%
By applying this formalism to a prototypical indirect materials bulk hBN, we computed the phonon-assisted radiative lifetime in a wide range of temperatures and found good agreement with the available experimental data. The resulting temperature dependence of the radiative lifetime is non-monotonic at low temperatures and monotonically decreases at higher temperatures above 150 K. This trend contrasts with the conventional \(T^{3/2}\) increase observed for direct exciton radiative lifetime~\cite{Las64,chen2019ab,im1997radiative}.
We explain the non-monotonic behavior at low temperatures with a simplified two-valley model, where a large difference in scattering rates between different valleys plays a critical role. At higher temperatures, the decrease in radiative lifetime is shown to be governed by the thermal occupation factors.
To capture the out-of-equilibrium dynamics of excitons, we derived and implemented a real-time quantum master equation in the Lindblad formalism. In the semiclassical limit, this reduces to the Boltzmann transport equation (BTE) with exciton-phonon scattering dynamics. Our simulations reveal ultrafast relaxation of excitons into finite-momentum valleys with information of relaxation and thermalization dynamics. 
This framework lays the groundwork for future extensions to study exciton coherence, dephasing, and exciton spin dynamics in realistic materials.

\section*{acknowledgement}
We thank very helpful discussions with Feliciano Giustino, Davide Sangalli, and Claudio Attaccalite. 
We acknowledge support from the Air Force Office of Scientific Research under AFOSR Award
No. FA9550-21-1-0087. 
This research used resources of the Scientific Data and Computing center, a component of the 
Computational Science Initiative, at Brookhaven National Laboratory under Contract No. DE-SC0012704,
the National Energy Research Scientific Computing Center (NERSC) a U.S. Department of Energy Office of Science User Facility operated under Contract No. DE-AC02-05CH11231. 
This work used the TACC Stampede3 system at the University of Texas at Austin through allocation PHY240212 from the Advanced Cyberinfrastructure Coordination Ecosystem: Services and Support (ACCESS) program~\cite{Boerner2023}, which is supported by US National Science Foundation grants No. 2138259, No. 2138286, No. 2138307, No. 2137603, and No. 2138296.

\bibliographystyle{apsrev4-2}
\bibliography{ref}

\end{document}